\documentclass[reprint,nofootinbib,amsmath,amssymb,aps,prmaterials]{revtex4-2}

\usepackage{graphicx}
\usepackage[dvipsnames]{xcolor}
\usepackage[colorlinks=true,
    linkcolor=blue,
    filecolor=blue,
    citecolor=blue,      
    urlcolor=blue,]{hyperref}

\usepackage{prettyref}
\newcommand{\pref}[1]{\prettyref{#1}}
\newrefformat{fig}{Fig.~\ref{#1}}
\newrefformat{tab}{Table~\ref{#1}}
\newrefformat{sec}{Sec.~\ref{#1}}
\newrefformat{subsec}{Sec.~\ref{#1}}
\newrefformat{app}{App.~\ref{#1}}
\newrefformat{eqn}{Eq.~(\ref{#1})}

\begin{document}

\title{Towards a unified first-principles-based description of VO$_2$ \\ using DFT+DMFT with bond-centered orbitals}

\author{Peter Mlkvik}
\author{Nicola A. Spaldin}
\author{Claude Ederer}
\email{claude.ederer@mat.ethz.ch}
\affiliation{Materials Theory, Department of Materials, ETH Z\"{u}rich, Wolfgang-Pauli-Strasse 27, 8093 Z\"{u}rich, Switzerland}

\date{\today}

\begin{abstract}
We present a combined density-functional theory and dynamical mean-field theory (DFT+DMFT) study of the full structural phase space of rutile-based vanadium dioxide (VO$_2$), including also the less studied M2 and T phases, using an unconventional bond-centered orbital basis. The use of bond-centered orbitals allows us to treat all main phases of VO$_2$, and the structural transitions between them, using one consistent approach with moderate computational cost and without pre-pattering of the structure into dimerized and undimerized V--V pairs. We obtain two distinct insulating states on the two different types of vanadium chains in the M2 phase, a singlet-insulator on the dimerized chains and a Mott-insulator on the zigzag-distorted chains, which, however, are strongly coupled in the M2 phase and thus the metal-insulator transition always occurs concomitantly for both types of sites. We also demonstrate that the M2 phase corresponds to a local energy minimum in the structural phase space of VO$_2$, the stability of which, apart from the internal structural distortion, depends crucially on the unit cell strain relative to the undistorted rutile phase. Our calculations further indicate that the symmetry-distinct triclinic T phase corresponds electronically to either an M1 or an M2-type insulator with an abrupt transition as a function of distortion. Finally, we disentangle the effect of the dimerization and zigzag distortions by constructing hypothetical structures that contain only one site type, finding that the zigzag distortion strongly favors emergence of the Mott-insulating state, both as function of distortion and on-site interaction. 
\end{abstract}

\maketitle


\section{Introduction}

\begin{figure}[!t]
	\centering
	\includegraphics[width=1\linewidth]{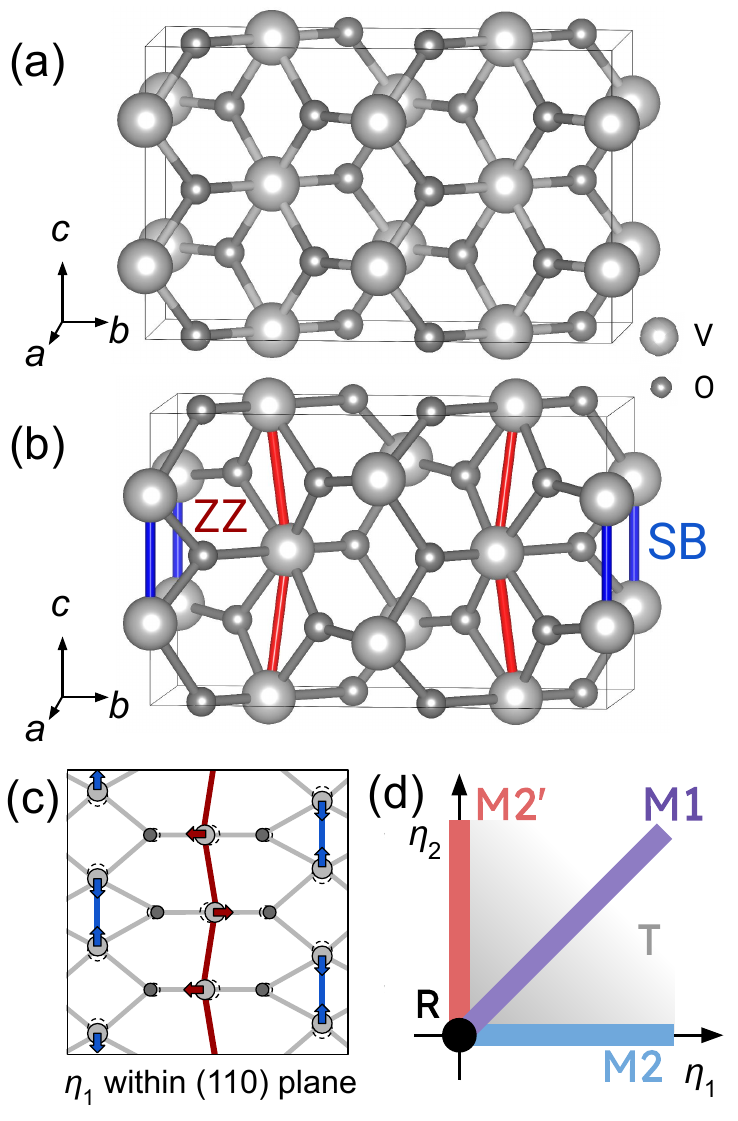}
	\caption{ (a) The R and (b) the M2 structures of VO$_2$ depicted in the unit cell used in this work. The short-bond pairs (SB) and zigzag-distorted chains (ZZ) are highlighted in blue and red, respectively. V (O) atoms shown in (dark) gray. (c) Distortion $\eta_1$ within the (110) plane [$\eta_2$ analogous in (1\=10) plane]. (d) The $(\eta_1, \eta_2)$ phase diagram, schematically indicating the R, M1, M2, and T phases. }
	\label{fig:m2-structure}
\end{figure}

Vanadium dioxide (VO$_2$) is a prototypical material exhibiting a metal-insulator transition (MIT), where the MIT is accompanied by a structural transition~\cite{Morin:1959}. At temperatures above $T_\text{MIT}$, VO$_2$ is metallic and exhibits the high-symmetry rutile (R) structure, which contains chains of equidistant nearest neighbor V atoms along the $c$ direction [see \pref{fig:m2-structure}(a)]. On cooling below $T_\text{MIT}$, VO$_2$ transforms into an insulating monoclinic, so-called M1 phase, which features a structural dimerization of the nearest neighbor V chains into alternating short bond (SB) and long bond (LB) V--V pairs, combined with a ``zigzag'' (ZZ) distortion of these chains due to additional displacements of the V atoms perpendicular to $c$~\cite{Eyert:2002a, Hiroi:2015, Pouget:2021}. Since the MIT occurs at $T_\text{MIT} \approx 340$\,K, i.e., not too far above room temperature, VO$_2$ is a promising test bed for many applications that attempt to exploit its MIT~\cite{Yang/Ko/Ramanathan:2011, Liu_et_al:2018, Yi_et_al:2018, Cui_et_al:2018, Shao_et_al:2018}. 

In principle, the insulating behavior of the M1 phase can be understood to result from a Peierls-like formation of electron pairs on the SB dimers, where the hybridization between the lowest-lying $a_{1g}$ orbitals of the nominal $d^1$ vanadium ions results in a strong bonding-antibonding splitting, while a simultaneous upwards energy shift of the higher-lying $e_g^\pi$ orbitals, ensures half-filling of the $a_{1g}$ states and thus leads to a band gap opening~\cite{Goodenough:1971}. However, this simple model has been called into question with proposals of Mott-Hubbard physics being important in VO$_2$~\cite{Zylbersztejn/Mott:1975, Qazilbash_et_al:2007, Huffman_et_al:2017, Najera_et_al:2017, Brito_et_al:2016, Brito_et_al:2017}, and thus the physics underlying the MIT in VO$_2$ has been the topic of a long-standing debate~\cite{Kosuge:1967, Wentzcovitch/Schulz/Allen:1994, Rice/Launois/Pouget:1994, Liebsch/Ishida/Bihlmayer:2005, Biermann_et_al:2005, Koethe_et_al:2006, Gatti_et_al:2007, Weber_et_al:2012, Belozerov_et_al:2012, Lee_et_al:2018, Weber_et_al:2020, Budai_et_al:2014}. Nowadays, the consensus is converging towards the idea that the formation of the insulating M1 phase in VO$_2$ is governed by a ``correlation-assisted'' Peierls mechanism~\cite{Biermann_et_al:2005, Tomczak/Aryasetiawan/Biermann:2008, Belozerov_et_al:2012, Mlkvik_et_al:2024}.

One of the arguments for the Mott-like nature of VO$_2$ stems from the fact that besides the R and M1 phases, VO$_2$ also forms another monoclinic, so-called M2, phase [\pref{fig:m2-structure}(b)]~\cite{Pouget_et_al:1974, Pouget_et_al:1975, Strelcov_et_al:2012, Quackenbush_et_al:2016}. In the M2 phase, the V chains are \emph{either} dimerized (with alternating SB and LB pairs along $c$) \textit{or} zigzag-distorted (with equidistant V atoms) [see \pref{fig:m2-structure}(b), with SB pairs and ZZ chains shown in blue and red, respectively]. While the M2 phase is found to be insulating, the insulating nature of the zigzag-distorted chains cannot be explained by the Peierls mechanism, with susceptibility measurements also indicating the presence of local magnetic moments indicative of a Mott-insulating state~\cite{Pouget_et_al:1974}. The study of the M2 phase can thus help better understand the general interplay between electronic correlation effects and structural distortions in the different phases of VO$_2$.

 The M2 phase was first observed in the 1970s~\cite{Marezio_et_al:1972, Pouget_et_al:1974, Pouget_et_al:1975} and the first phase diagrams containing R, M1, and M2 phases were constructed~\cite{Pouget_et_al:1975}. A strain within the [110] plane was shown to yield the M2 phase in an intermediate temperature regime, leading to an R to M2 transition. On further lowering of the temperature, the M2 phase then transforms into the M1 phase through an intermediate triclinic, so-called T phase~\cite{Ghedira_et_al:1977}. Interestingly, light Cr or Al doping also yields the M2 phase~\cite{Marezio_et_al:1972, Pouget_et_al:1974, Ghedira_et_al:1977}. Thus, since the 1970s, there have been many refinements to the VO$_2$ phase diagram, including not only R and M1, but also the M2 and T phases~\cite{Strelcov_et_al:2012, Mandal_et_al:2025, Zhang/Chou/Lauhon:2009, Cao_et_al:2010, Atkin_et_al:2012, Okimura/Watanabe/Sakai:2012, Quackenbush_et_al:2016}.

Although many first principles-based electronic structure calculations exist for the R and M1 phases, see, e.g., Refs.~\cite{Eyert:2002a, Weber_et_al:2012, Gatti_et_al:2007, Weber_et_al:2020, Eyert:2011, Grau-Crespo/Wang/Schwingenschlogl:2012, Kylanpaa_et_al:2017, Zheng/Wagner:2015, Biermann_et_al:2005, Tomczak/Biermann:2007, Tomczak/Biermann:2009}, theoretical studies of the M2 and T phases, and also of the structural transformations between the different phases, are rather sparse. Initially described as a simple Heisenberg spin chain~\cite{Pouget_et_al:1974}, the M2 phase is often modeled using spin-polarized density-functional theory (DFT) calculations with antiferromagnetically aligned V cations within the zigzag distorted chains~\cite{Eyert:2002a, Eyert:2011, Yuan_et_al:2012, Zhang_et_al:2024}. However, Brito~\textit{et al.}~\cite{Brito_et_al:2016, Brito_et_al:2017} used DFT plus dynamical mean-field theory (DMFT) to show that, while the dimerized chains form a singlet insulator similar to the M1 phase, the zigzag-distorted chains indeed form a Mott insulator. These authors also compared paramagnetic to antiferromagnetic calculations and concluded that the effects of magnetic order on the band gap are weak relative to the Mott effects. 

Specifically, the work of Brito~\textit{et al.}~\cite{Brito_et_al:2016, Brito_et_al:2017} employed a single-site DMFT treatment for the undimerized, and a two-site cluster DMFT treatment for the dimerized V atoms, hence treating the different types of V chain on different footing. Similarly, the earlier seminal work of Biermann~\textit{et al.}~\cite{Biermann_et_al:2005} used DFT plus single-site DMFT for the description of the R phase, and two site cluster DMFT for the M1 phase. Thus, while these different variants of the DFT+DMFT method have been successfully used to describe different phases of VO$_2$, and are capable of describing both the weakly correlated and strongly correlated limits equally well, the required ``pre-patterning'' into dimerized and undimerized V pairs represents an obstacle for further studies. Specifically, in order to compare relative energies of the different phases or continuously transform one structural variant into another, it is desirable to use a universal computational framework that allows to treat all phases on equal footing.     

We have addressed this problem in a preceding work~\cite{Mlkvik_et_al:2024} by constructing a set of ``bond-centered'' orbitals, i.e., centered in between neighboring V atoms along the $c$-axis chains, which are then used to define the correlated subspace for a single-site DFT+DMFT calculation. By constructing the correlated orbitals on bond-centers between V atoms, this approach allows us to capture the singlet-insulating state realized in M1 VO$_2$, essentially replicating the results of DFT + cluster DMFT calculations at significantly lower computational cost. Simultaneously, this method also allows us to describe the undimerized R phase, or a potential competing Mott-insulating state, and thus, in principle, to explore the full structural phase space of VO$_2$ using a consistent computational framework (and without pre-patterning of potentially dimerized V--V pairs).

In the present work, we extend the application of the bond-centered DFT+DMFT approach also to the M2 phase of VO$_2$, for which we obtain a distinct Mott and Peierls behavior of the two different types of V chains, zigzag and dimerized, respectively, in agreement with the results of Brito~\textit{et al.}~\cite{Brito_et_al:2016, Brito_et_al:2017} (see \pref{sec:characterizing}). In addition, we characterize VO$_2$ throughout the full structural phase space (as defined in \pref{sec:eta}), extending our previous treatment of the R--M1 distortion from Ref.~\cite{Mlkvik_et_al:2024} towards the R--M2 and M1--T--M2 structural distortions and investigate the resulting effect on the electronic properties in the various metallic and insulating regimes. Finally, in \pref{sec:disentangling}, we separately analyze the effects of different components of the structural distortions present in VO$_2$, namely the unit cell strain (relative to the R phase) and the internal V displacements, also focusing on the separate roles of the structural dimerization and zigzag distortion.


\section{Definition of the structural phase space}
\label{sec:eta}

The distortion pattern in the M2 structure corresponding to either dimerized or zigzag-distorted vanadium-vanadium chains can also be recast as a set of collective displacements, $\eta$, of vanadium atoms within a (110) plane [see \pref{fig:m2-structure}(c)] or, equivalently, within a (1\=10) plane~\cite{Brews:1970, Tselev_et_al:2010, Davenport_et_al:2021}. Within these planes, the displacements of the V atoms in each zigzag-distorted chain are coupled to the displacements that lead to the formation of alternating SB and LB pairs in the neighboring chain, as indicated by the arrows in \pref{fig:m2-structure}(c). The coupling can be thought of as being mediated by the oxygen anion shared between the corresponding oxygen octahedra (apical relative to the zigzag distorted V, and equatorial with respect to the dimerizing V).

In the M2 phase, these distorted planes are stacked along the direction perpendicular to these (110)-type planes, while the M1 phase is formed by an orthogonal pair of such displacements equal in strength, $\eta_1 = \eta_2$, corresponding to displacements within (110) and (1\=10) planes, respectively, while for $\eta_1 \neq \eta_2$, the triclinic phase, T, is obtained, bridging the M1 and M2 structures~\cite{Ghedira_et_al:1977}. In \pref{fig:m2-structure}(d), we show a section of the full $(\eta_1, \eta_2)$ phase space, where we indicate that $\eta_1$ [within a (110) plane] and $\eta_2$ [within a (1\=10) plane] individually lead to distinct domains of the M2 structure and when they are equal they yield the M1 structure. More generally, the $(\eta_1, \eta_2)$ distortion patterns are a subset of the full four-component order parameter governing the structural distortions in VO$_2$~\cite{Brews:1970, Tselev_et_al:2010, Davenport_et_al:2021}, where the other components allow for the formation of dimerizing and zigzag distortions on the other respective chains.

\section{Computational Method}

We conduct DFT+DMFT calculations with a bond-centered orbital basis, following Ref.~\cite{Mlkvik_et_al:2024}. This choice of basis is motivated by physical intuition about the molecular-like nature of the insulating state in M1 VO$_2$. However, the bond-centered orbitals are centered between \emph{every} neighboring V--V pair along $c$ and hence the construction is agnostic to positive or negative direction of dimer formation, i.e., there is no \textit{a priori} pairing of vanadium atoms. We construct Wannier functions that have bonding character on the V--V pairs of interest and host antibonding tails on the neighboring pairs.

\begin{figure}[!t]
	\centering
	\includegraphics[width=1\linewidth]{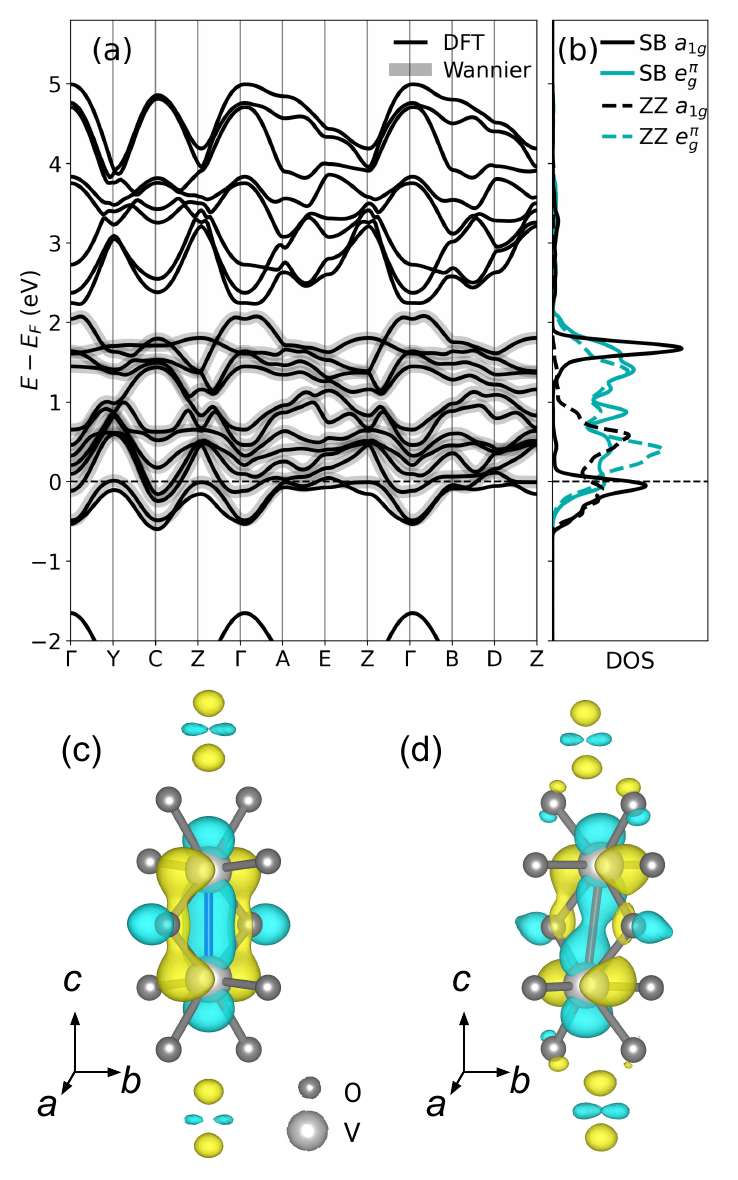}
	\caption{(a) DFT band structure of the M2 VO$_2$ shown in black together with the bands recalculated within the Wannier basis in gray. (b) Density of states (DOS) projected on V $a_{1g}$ ($e_g^\pi$) orbital character plotted in black (cyan). Orbitals on the SB (ZZ) sites shown as full (dashed) lines. (c, d) Bond-centered orbitals corresponding to the $a_{1g}$ orbitals on the (c) SB and (d) ZZ sites. Yellow (cyan) colors indicate the positive (negative) phase of the orbitals. }
	\label{fig:m2-bandstructure}
\end{figure}

Practically, we first transform the V-$t_{2g}$-dominated set of bands around the Fermi level [indicated as Wannier bands in \pref{fig:m2-bandstructure}(a)] into a set of atom-centered Wannier functions, which we then transform into a bond-centered basis using a $\textbf{k}$-dependent transformation matrix~\cite{Mlkvik_et_al:2024},
\begin{equation} 
	\label{eq_rotmat}
	U(\textbf{k}) =  \frac{1}{\sqrt{2}}
	\begin{pmatrix}
		e^{i(\pi/4 - k_z c/4)} & e^{i(-\pi/4 + k_z c/4)}\\
		e^{i(-\pi/4 - 3 k_z c/4)} & e^{i(\pi/4 - k_z c/4)}\\
	\end{pmatrix},
\end{equation}
which always combines two atom-centered Wannier functions with equivalent orbital character corresponding to neighboring V sites along $c$. The so-defined orbitals are centered in the middle of the nearest neighbor V--V pairs along the $c$ direction [see SB and ZZ bond-centered functions in the M2 phase in \pref{fig:m2-bandstructure}(c, d), respectively]. We construct a bond-centered orbital from each of the $a_{1g}$ and $e_g^\pi$ orbitals, with the functions spanning the same correlated subspace as the more conventional atom-centered orbitals. For more details, we refer to Ref.~\cite{Mlkvik_et_al:2024}.

We use a periodically repeated unit cell that corresponds to an approximately orthorhombic $1\times2\times2$ supercell of the underlying R unit cell that can describe both M1 and M2 structures and contains eight VO$_2$ formula units. Unless otherwise specified, we use the experimental lattice parameters for the respective R structure~\cite{McWhan_et_al:1974}, M1 structure~\cite{Longo_et_al:1970}, and M2 structure~\cite{Marezio_et_al:1972} summarized in \pref{tab:latticeparams}. When we perform structural variation, unless otherwise specified, we interpolate between these experimental lattice parameters, and vary both vanadium and oxygen internal coordinates accordingly.

\setlength{\tabcolsep}{6pt}
\begin{table}[!t]
    \centering
    \begin{tabular}{l c c c c c c}
    \hline\hline
           & $a$\,(Å) & $b$\,(Å) & $c$\,(Å) & $\alpha$\,($^\circ$) & $\beta$\,($^\circ$) & $\gamma$\,($^\circ$)  \\ \hline
         R & 4.55 & 9.11 & 5.70 & 90.0 & 90.0 & 90.0 \\
         M1 & 4.53 & 9.08 & 5.75 & 90.0 & 90.4 & 90.0 \\
         M2 & 4.53 & 9.07 & 5.80 & 90.0 & 90.0 & 91.9  \\
         \hline\hline
    \end{tabular}
    \caption{Lattice parameters of our unit cell corresponding to the experimental lattice parameters of the R~\cite{McWhan_et_al:1974}, M1~\cite{Longo_et_al:1970}, and M2~\cite{Marezio_et_al:1972} structures used in this work. Lattice parameters given in Å and angles in degrees. }
    \label{tab:latticeparams}
\end{table}

We perform DFT calculations using the \textsc{quantum espresso} (v7.3) package~\cite{Giannozzi_et_al:2009, Giannozzi_et_al:2017} within the generalized gradient approximation, using the Perdew-Burke-Ernzerhof~\cite{Perdew/Burke/Ernzerhof:1996} exchange-correlation functional. We use ultrasoft pseudopotentials from the GBRV library~\cite{Garrity_et_al:2014}, including the semicore 3$s$ and 3$p$ states of the vanadium atoms in the valence manifold. We use a wavefunction plane-wave cutoff of 70\,Ry, and 12$\times$70\,Ry for the charge density. We use a $\Gamma$-centered $5\times6\times8$ $k$-point mesh. We converge the total energies to $5\times10^{-8}$\,eV.

We use \textsc{wannier90} (v3.1.0)~\cite{Mostofi_et_al:2014, Pizzi_et_al:2020} to construct the Wannier functions. We use the vanadium $t_{2g}$-dominated manifold from around $-0.5$\,eV to $2$\,eV isolated from the rest of the bands [see the highlighted bands in \pref{fig:m2-bandstructure}(a)].

We perform both one-shot and charge self-consistent DFT+DMFT calculations~\cite{Georges_et_al:1996, Kotliar_et_al:2006, Lechermann_et_al:2006, Beck_et_al:2022} using \textsc{solid\_dmft}~\cite{Merkel_et_al:2022} within the TRIQS (v3.3.0) software suite~\cite{Parcollet_et_al:2015}. We parameterize the local interaction on each bond-center by a Hubbard-Kanamori Hamiltonian including the spin-flip and pair-hopping terms~\cite{Kanamori:1963, Vaugier/Jiang/Biermann:2012, Georges/Medici/Mravlje:2013} (shown to be a good approximation for the bond-centered orbitals, see Ref.~\cite{Mlkvik_et_al:2024}). We first vary the interaction parameters over a realistic range to explore the phase diagram at a fixed structure. Then, when imposing structural distortions, we keep the interaction parameters fixed close to the values calculated via the constrained random phase approximation (cRPA) for the bond-centered orbitals in the R phase, $(U_\text{cRPA},J_\text{cRPA})=(1.35,0.19)$\,eV~\cite{Mlkvik_et_al:2024}, which are very similar to those of the M1 phase. The justification for fixing the interaction parameters stems from the fact that we have previously found only very small differences in calculated parameters between the R and M1 phases~\cite{Mlkvik_et_al:2024}, and we thus expect this to hold for the other structural variants as well.

We solve the DMFT impurity problem using CT-HYB~\cite{Seth_et_al:2016}, a continuous-time quantum Monte Carlo solver~\cite{Werner/Millis:2006, Gull_et_al:2011}, setting the inverse electronic temperature, $\beta = 40\,\text{eV}^{-1}$, which corresponds to approximately room temperature. We represent the Green's function using the Legendre polynomial expansion, considering the first 30 terms. The DMFT problem is solved self-consistently, and converged until orbital occupation differences between consecutive steps are smaller than $10^{-2}$ electrons. We employ the fully localized limit double counting correction~\cite{Anisimov_et_al:1993}, and converge the total energy to $5\times10^{-2}$\,eV. We ensure a paramagnetic solution by averaging over both spin channels.

From the local Green's function, we obtain local occupations as well as the spectral weight at zero frequency, \mbox{$\bar{A}(\omega=0)=-(\beta/\pi)$Tr$G(\tau=\beta/2)$}. We estimate the quasiparticle weight, $Z = [1-\partial$Im$\Sigma(i\omega)/\partial(i\omega)]^{-1}$, from a linear fit to the imaginary part of the local self energy at the smallest two Matsubara frequencies , interpolating to $i\omega=0$. Finally, we use the maximum entropy method to obtain the $k$-summed spectral functions on the real-frequency axis~\cite{Jarrell/Gubernatis:1996, Kraberger_et_al:2017, Bergeron/Tremblay:2016}.


\section{Characterizing the M2 phase}
\label{sec:characterizing}

We start by investigating whether the bond-centered basis of the correlated subspace is appropriate for the study of the M2 phase. We analyze different regimes observed for different interaction parameters, identifying the one representing the M2 phase, and we also consider a structural pathway that connects the M2 phase to the previously calculated R and M1 phases of VO$_2$.

\subsection{M2 phase diagram in \textit{U} and \textit{J}}
\label{subsec:m2diag}

We perform one-shot DFT+DMFT calculations in the bond-centered basis for a wide range of interaction parameters to explore and characterize the different possible phases obtained for the fixed M2 structure. In particular, we vary the $U$ and $J$ parameters around their calculated cRPA values in the bond-centered basis, $(U_\text{cRPA},J_\text{cRPA})=(1.35,0.19)$\,eV \cite{Mlkvik_et_al:2024}.

\begin{figure*}
	\centering
	\includegraphics[width=1\textwidth]{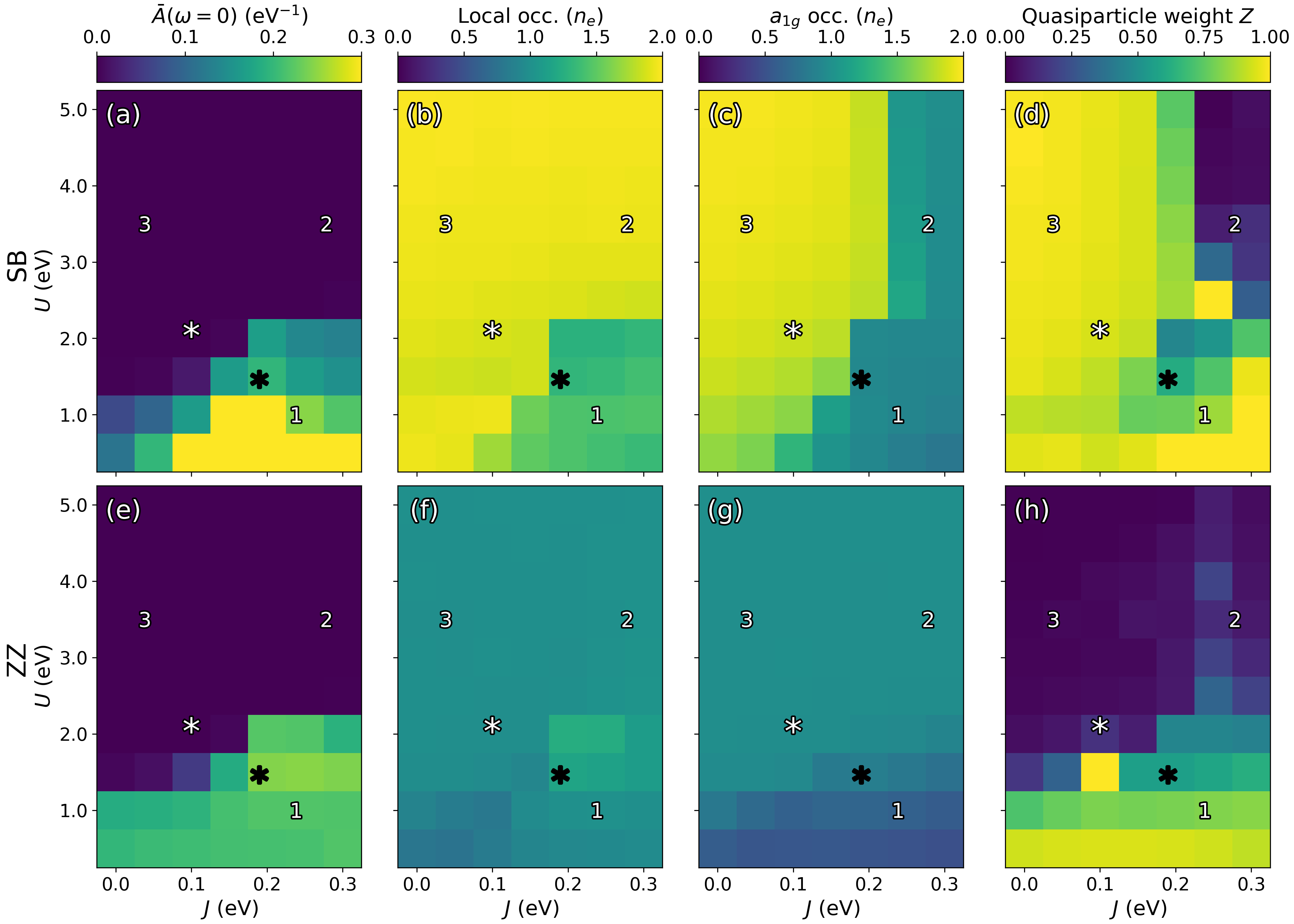}
	\caption{Different local observables on the M2 (a-d) SB and (e-h) ZZ sites obtained within bond-centered DFT+DMFT as a function of $U$ and $J$: (a, e) spectral weight at zero frequency, $A(\omega=0)$, (b, f)  total local occupation, (c, g) occupation of the $a_{1g}$ orbital, and (d, h) the $a_{1g}$ quasiparticle weight $Z$. Labels (1-3) indicate the different regimes discussed in the text. Black-filled stars indicate the cRPA values of $(U, J)$ obtained for the R phase in \cite{Mlkvik_et_al:2024}, while white-filled stars indicate the values used in subsequent calculations.}
	\label{fig:m2-ujdiag}
\end{figure*}

As discussed above, the M2 structure, due to the difference in neighboring V--V chains, hosts three distinct bond-centered sites -- short-bond (SB), long-bond (LB), and the zigzag-distorted (ZZ). In \pref{fig:m2-ujdiag}(a-h), we only show results for the SB and ZZ sites, since the corresponding properties of the LB sites can usually be inferred from those of the SB and ZZ sites.

From the local spectral weight at zero frequency shown in \pref{fig:m2-ujdiag}(a, e) one can see that for $U \lesssim 2$\,eV the system is metallic, while for $U \gtrsim 2$\,eV we obtain an insulator, indicated by $\bar{A}(0)\approx 0$. There is also a weak $J$ dependence with larger $J$ favoring the metallic state. Notably, the transition between metal and insulator occurs simultaneously on all bond-centers (also on the LB sites not shown here), i.e., we observe no $(U,J)$ regime with site-selective metallicity, in spite of the different character of the insulating state on the SB/LB and ZZ sites, which we now analyze further.

The different character of the insulating state on the SB (and LB) and the ZZ sites can for example be inferred from the total local occupation on these sites, shown in \pref{fig:m2-ujdiag}(b, f). While, in the insulating regime, the ZZ sites are occupied by exactly one electron, the SB sites are doubly occupied with two electrons (and since the correlated subspace contains one electron per V atom, this leaves zero electrons for the LB sites). Further inspection of the orbital-resolved $a_{1g}$ occupations shown in \pref{fig:m2-ujdiag}(c, g) reveals that one can in fact distinguish two different insulating regimes on the SB sites. For $J \lesssim 0.25$\,eV, both electrons on the SB sites occupy the lowest lying $a_{1g}$ orbital, while, for larger $J$, one of these electrons is transferred to the higher-lying $e_g^\pi$ states. Thus, for $J\lesssim 0.25$\,eV, the insulating state on the SB sites represents a singlet-insulator, similar to that observed in the insulating M1 phase~\cite{Mlkvik_et_al:2024}, while, for larger Hund's interaction $J$, the singlet is transformed into a Mott-like triplet, with two electrons distributed over different orbitals and an emerging local magnetic moment on the SB site. The ZZ (and LB) sites are not affected by this low-spin to high-spin transition on the SB sites.

The transition on the SB sites from an essentially uncorrelated singlet insulator to a correlated Mott insulator around $J \approx 0.25$\,eV can also be seen from the local quasiparticle weights, $Z$, shown in \pref{fig:m2-ujdiag}(d, h). For $J \lesssim 0.25$\,eV (and in the metallic regime), the quasiparticle weight stays close to $Z=1$, indicating a weakly correlated state, whereas for $J\gtrsim 0.25$\,eV, the quasiparticle weight drops to zero, indicative of a Mott insulator. Furthermore, one can see that on the ZZ sites the quasiparticle weight essentially vanishes across the whole insulating region for $U\gtrsim 2$\,eV, establishing the Mott-insulating character of the ZZ sites.

Thus, we can distinguish three different regimes in the $(U,J)$ phase diagrams shown in \pref{fig:m2-ujdiag}, indicated by labels 1, 2, and 3: (1) a metallic regime for small $U$, (2) an insulating regime with a one-electron Mott-insulating state on the ZZ sites and a two-electron Mott-triplet on the SB sites for $U\gtrsim 2$\,eV and large $J$ (combined with a trivial unoccupied band-insulator on the LB sites), and (3) a singlet-insulator on the SB sites coexisting with the Mott state on the ZZ sites. Thus, regime 3 can be identified as representative for the actual M2 phase of VO$_2$, by analogy with the results obtained in the DFT + cluster DMFT study of Brito {\it et al.}~\cite{Brito_et_al:2016}, which also found a singlet-insulating state in the dimerized SB/LB chains and a singly occupied Mott state on the ZZ chains. 

\begin{figure}
	\centering
	\includegraphics[width=1\linewidth]{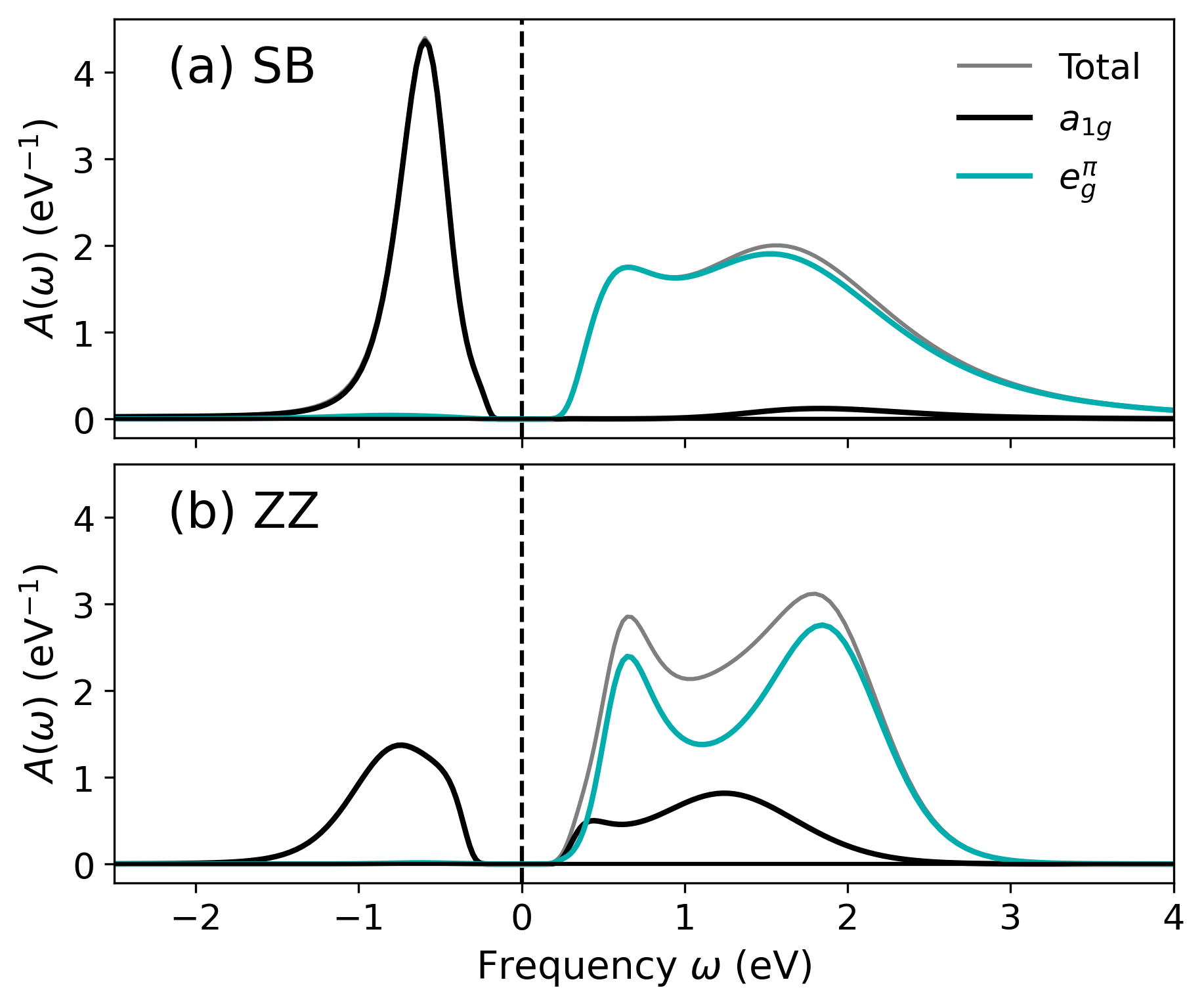}
	\caption{Local spectral functions in M2 VO$_2$ at $(U, J)=(2.0,0.1)$\,eV for the (a) SB and (b) ZZ sites. Black (cyan) lines show the $a_{1g}$ ($e_g^\pi$) bands, gray line shows the total per impurity. Dashed line indicates the Fermi level.}
	\label{fig:m2-spectral}
\end{figure}

The different character of the insulating state on the SB and ZZ sites in regime 3 can also be seen from the corresponding local orbital-resolved spectral functions in \pref{fig:m2-spectral}. On the SB sites, the $a_{1g}$ states are completely filled (with two electrons), while on the ZZ sites the $a_{1g}$ state are only half-filled and exhibit a Mott gap. On both sites, the higher lying $e_g^\pi$ states remain empty.

The interaction parameters obtained from cRPA calculations in our previous work, Ref.~\cite{Mlkvik_et_al:2024}, and indicated by the black-filled stars in \pref{fig:m2-ujdiag}, are rather close to the boundary between all three regimes. We take this as confirmation that the bond-centered basis indeed allows for a realistic description of the M2 phase of VO$_2$, since it is well established that, while cRPA calculations can provide realistic estimates of the strength of the screened interaction, the corresponding values often need small adjustments to obtain optimal effective interaction parameters to be used in realistic calculations~\cite{Honerkamp_et_al:2018, Pauli_et_al:2025a, Carta/Panda/Ederer:2026, vanLoon_et_al:2021, Scott/Booth:2024a, Merkel/Ederer:2024}. In the following, we will therefore use values $(U,J)=(2.0, 0.1)$\,eV, indicated by the white-filled stars in \pref{fig:m2-ujdiag}, which place the system firmly into the realistic regime 3. These values also mirror the $(U,J)$ values used to study the M1 and R phases in Ref.~\cite{Mlkvik_et_al:2024}, and thus allow us to describe all phases of VO$_2$, and thus explore the complete $(\eta_1,\eta_2)$ structural phase space, using a consistent choice of interaction parameters.

\subsection{Variation of M2 structural distortion}
\label{subsec:m2dist}

Next, we perform charge self-consistent DFT+DMFT calculations where we systematically vary the structural distortion leading from the R and M1 to the M2 phase, studying the behavior across the MIT with increasing structural distortion. We trace both the R, $(0,0)$, to M2, $(\eta,0)$, and the M2, $(\eta, 0)$, to M1, $(\eta, \eta)$, transition [see \pref{fig:m2-structure}(d)]. As discussed at the end of \pref{subsec:m2diag}, we use $(U,J)=(2.0,0.1)$\,eV.

\begin{figure}[!t]
	\centering
	\includegraphics[width=1\linewidth]{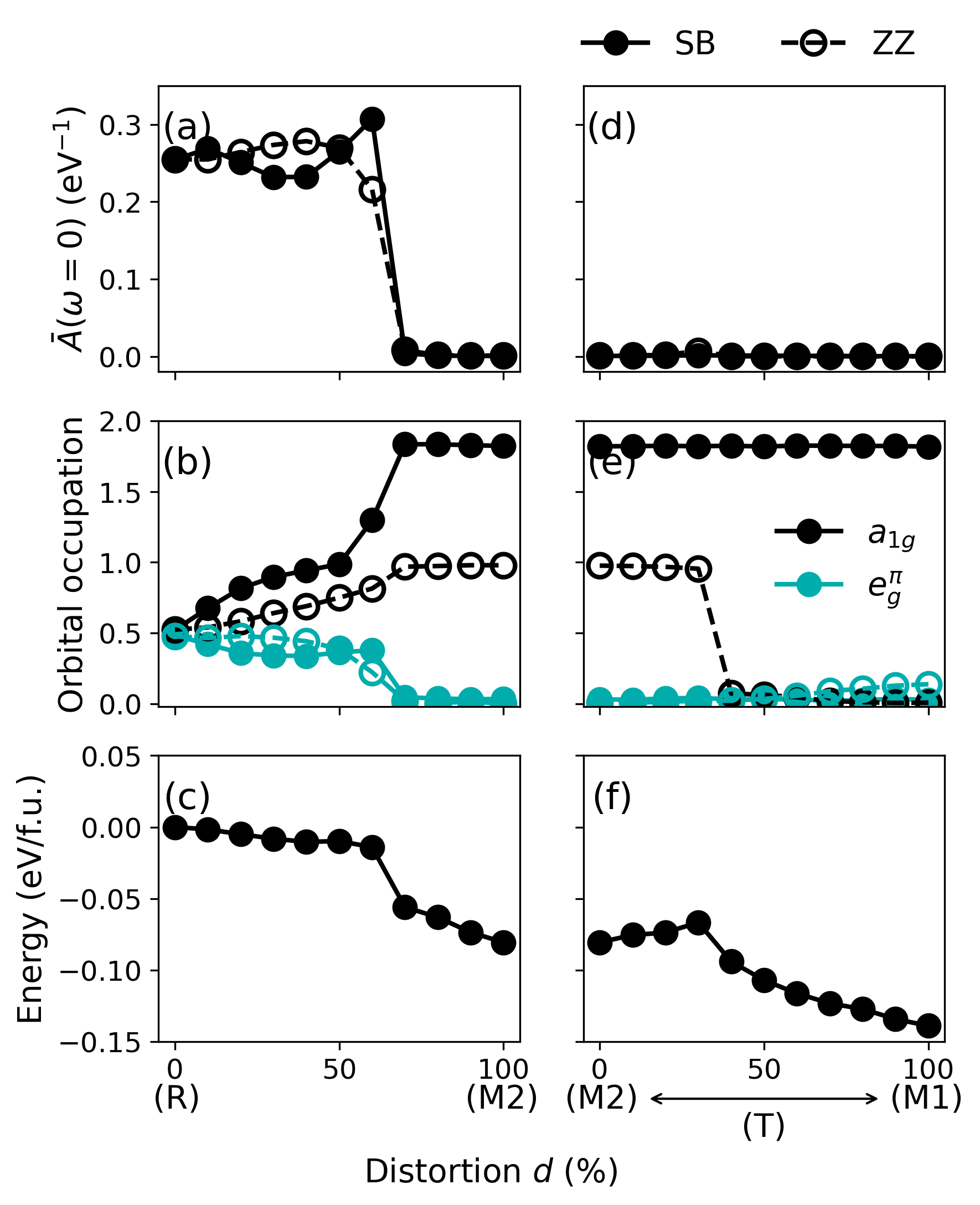}
	\caption{ Selected observables calculated as function of structural distortion interpolating between (a-c) R and M2, and (d-f) M2 and M1 (through T) structures for the SB (solid line, filled circles) and ZZ (dashed line, open circles) sites. (a, d) Spectral weight at zero frequency. (b, e) Orbital occupation of the $a_{1g}$ (black) and the two $e_g^\pi$ (cyan) orbitals. (c, f) Total energy relative to the R phase.}
	\label{fig:m2-dist}
\end{figure}

In \pref{fig:m2-dist}, we show the results for both the SB (full lines, filled markers) and ZZ (dashed lines, empty markers) sites, indicating the spectral weight at zero frequency [\pref{fig:m2-dist}(a, d)], orbital occupation [\pref{fig:m2-dist}(b, e)], and the total energy [\pref{fig:m2-dist}(c, f)], all as a function of the structural distortion interpolated between the two end members. Note that for the M2--M1 path, the two ZZ sites become inequivalent and only the one that evolves into the LB site in the M1 limit is shown.

First focusing on the R--M2 transition~[\pref{fig:m2-dist}(a-c)], at small distortions away from the R structure, VO$_2$ remains metallic on both the SB and ZZ sites~[\pref{fig:m2-dist}(a)]. The occupation of the $a_{1g}$ and $e_g^\pi$ orbitals changes, leading to a gradual depletion of the two $e_{g}^\pi$ orbitals~[\pref{fig:m2-dist}(b)]. As we move away from the R phase, we also see an energy lowering, indicating an unstable nature of the R phase~[\pref{fig:m2-dist}(c)]. 

At around 60\% of distortion, we observe an abrupt MIT occurring simultaneously on both SB and ZZ sites~[\pref{fig:m2-dist}(a)]. The electrons localize in the $a_{1g}$ orbitals, now completely depleting the $e_g^\pi$ orbitals on both chains, leading to an occupation of two and one on the SB and ZZ sites, respectively~[\pref{fig:m2-dist}(b)], corresponding to zero electrons on the LB sites. The MIT is also accompanied by a further energy lowering~[\pref{fig:m2-dist}(c)]. When extrapolating the structural distortion beyond 100\% (not shown), we find an energy minimum corresponding to the M2 phase at around 120\% distortion, while for even larger distortion, the energy rises again, establishing the M2 phase as a locally stable phase with respect to this distortion mode.

Looking at the M2--T--M1 transition~[\pref{fig:m2-dist}(d-f)], close to the M2 structure, the system remains in an M2-like phase even though its structural symmetry is now that of the T phase. All bond centered sites remain insulating~[\pref{fig:m2-dist}(d)], with two electrons occupying the SB sites and one electron occupying the ZZ bonds~[\pref{fig:m2-dist}(e)]. We note that these occupations are ``locked'' to integer values in the respective insulating regimes. We further see that the energy increases with distortion and thus the M2 phase is also locally stable with respect to this distortion mode~[\pref{fig:m2-dist}(f)].

At around 40\% distortion, an abrupt transition occurs. Although the system remains insulating throughout~[\pref{fig:m2-dist}(d)], the occupations show that the nature of the insulator changes. The SB sites remain doubly occupied, while the ZZ site shown here becomes completely empty~[\pref{fig:m2-dist}(e)], effectively turning into an M1-like LB site. Conversely, the other ZZ site (not shown here) gains one electron, and becomes electronically equivalent to the doubly occupied SB site. This electron rearrangement is accompanied by an energy lowering for further increasing distortion, with the potentially global energy minimum close to the experimental M1 structure~[\pref{fig:m2-dist}(f)], while for distortions larger than 130\%, the energy rises again. The Mott-insulating ZZ chains have thus changed into chains of alternating empty and doubly occupied singlet-insulating sites, completing the path through the phase diagram.

In summary, we find an MIT between the R and M2 phase where SB and ZZ sites become insulating simultaneously, giving rise to a local energy minimum corresponding to the M2 phase. Upon distorting through the T phase into the M1 phase, the M2 phase remains locally stable before we observe an abrupt transition into the M1 phase corresponding to the global energy minimum. Notably, although structurally the whole range of distortion between the M2 and M1 phases is characterized as T phase, electronically, we observe either an M2-like insulating state or an M1-like insulating state. In experiment, only one clear transition has been found, a first-order transition between the M2 and T phase, while the T to M1 phase transition is classified as a continuous crossover~\cite{Strelcov_et_al:2012, Huffman_et_al:2017}. Our results appear to be consistent with this observation, indicating that the T phase might essentially correspond to a distorted M1 phase. To the best of our knowledge, this is the first first-principles characterization of the T phase, apart from a calculation of the optical conductivity spectra in Ref.~\cite{Huffman_et_al:2017}.

\section{Disentangling unit cell and internal distortion effects}
\label{sec:disentangling}

In the previous section, we varied both the internal distortion within the $(\eta_1, \eta_2)$ subspace and simultaneously interpolated between the experimental cell parameters of the R, M1, and M2 VO$_2$ phases. In this section, we aim to disentangle the influence of the different components of the structural distortion by, first, addressing the effect of the lattice parameter changes with the internal atomic coordinates fixed to those of the M2 structure, and, second, studying hypothetical structures featuring only dimerized or zigzag-distorted chains.

\subsection{Unit cell distortion effects}
\label{subsec:m2-straineffects}

\begin{figure}[!t]
	\centering
	\includegraphics[width=1\linewidth]{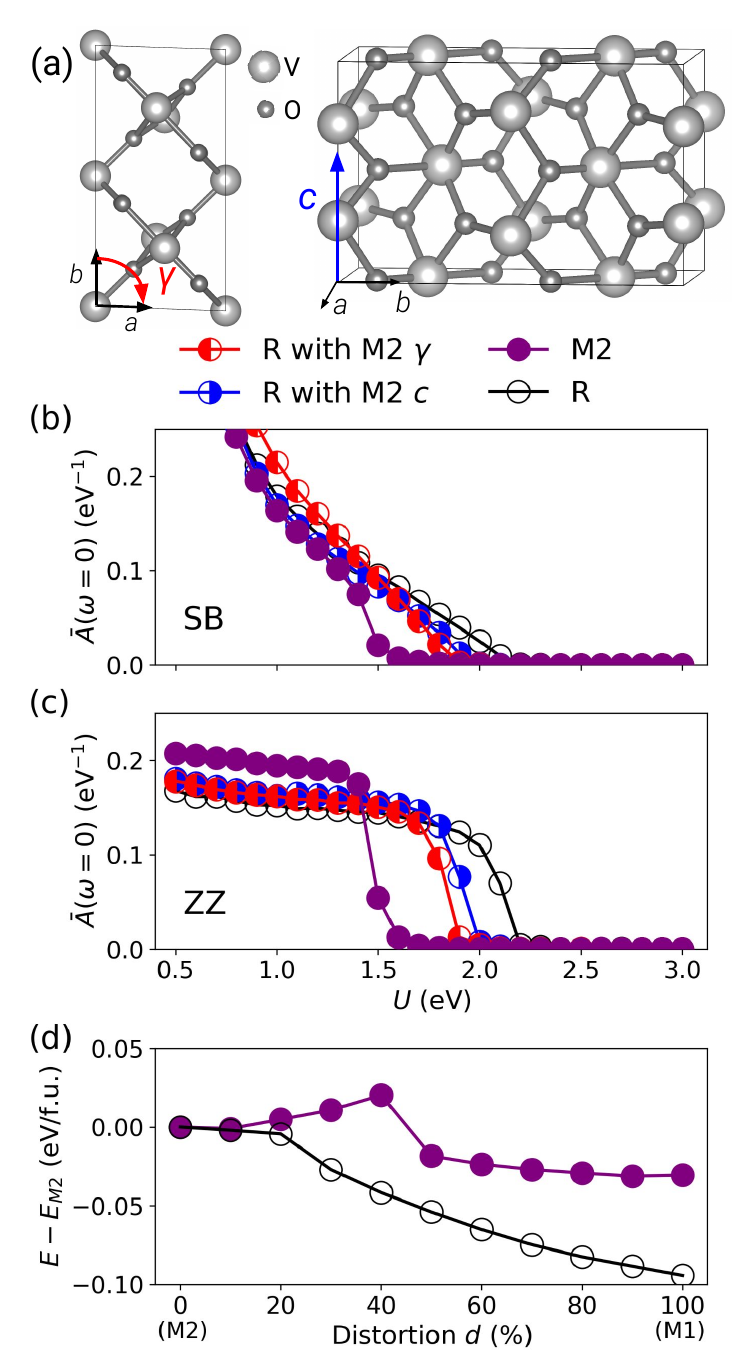}
	\caption{ (a) The M2 structure of VO$_2$, with the monoclinic angle $\gamma$ and the $c$ axis strain highlighted by red and blue arrows, respectively. V (O) atoms are shown in light (dark) gray. (b, c) Local spectral weight at zero frequency as a function of $U$ for the SB and ZZ bonds, respectively, with the M2 internal structural distortion embedded in the R unit cell (empty markers), M2 unit cell (purple markers), R unit cell with the M2 $\gamma$ angle (half-filled red markers), and the R unit cell with the M2 $c$ lattice parameter (half-filled blue markers). (d) Energy as a function of internal distortion between the M2 and M1 structures embedded in the R (empty markers), and M2 (purple markers) unit cells at $(U, J)=(2.0,0.1)$\,eV.}
	\label{fig:m2-strain}
\end{figure}

Inspecting the difference between the high-symmetry reference R and lower-symmetry distorted M2 unit cells (\pref{tab:latticeparams}), we note that the most notable changes are in the $\gamma$ angle and the expansion of the $c$ lattice parameter in the M2 phase [schematically indicated by red and blue arrows, respectively, in \pref{fig:m2-strain}(a)]. The differences in the $a$ and $b$ lattice parameters between R and M2 unit cells are comparatively small. In the following, we therefore study how the changes of $\gamma$ and $c$ affect the transition between the R and M2 phases.

We first perform one-shot DFT+DMFT calculations for varying $U$ and fixed $J=0.1$\,eV, i.e., corresponding to a vertical cut across the phase diagrams shown in \pref{fig:m2-ujdiag}. In \pref{fig:m2-strain}(b, c), we show the spectral weight at zero frequency on the SB and ZZ sites, obtained using the lattice parameters of the M2 unit cell, the R unit cell, the R unit cell including only the value of $\gamma$ corresponding to the M2 phase (``R with M2 $\gamma$''), and the R unit cell including only the change in $c$ (``R with M2 $c$''). In all cases the internal relative atomic coordinates are fixed to those of the M2 structure. We observe slightly different behaviors with increasing $U$ on the two different sites, indicating their distinct insulating states as discussed previously. The transition to the singlet insulating state on the SB sites features a gradual decrease in spectral weight with $U$. On the other hand, the ZZ sites exhibit a more abrupt transition to the Mott-insulting state at a critical $U$ value. However, in spite of their distinct character, the MIT happens concomitantly on both the ZZ and SB sites in all cases, similarly to what was observed in the previous sections.

One can further observe a notable effect of the different lattice parameters on the size of $U$ required to trigger the MIT, which is reduced by over 30\,\% (from 2.2\,eV to 1.5\,eV) when using the M2 lattice parameters (purple markers) compared to those of the R cell (empty markers). When only the changes in either $c$ (half-filled blue markers) or $\gamma$ (half-filled red markers) are included individually, the reduction of the critical $U$ is in-between, indicating that both have a comparable and cooperative effect on the MIT.

Finally, we also investigate how the changes of the unit cell parameters affect the energetic stability of the M2 phase using charge self-consistent DFT+DMFT calculations at $(U, J)=(2.0,0.1)$\,eV. In \pref{fig:m2-strain}(d), we show the total energy as a function of the M1--M2 internal distortion, once calculated with lattice parameters fixed to the M2 unit cell (purple markers), and once calculated with lattice parameters fixed to those of the R unit cell (empty markers). One can see that the change in the lattice parameters has an important effect on the energetic stabilization of the M2 phase. While the calculations using the fixed M2 lattice parameters exhibit a local energy minimum around the M2 structure, similar to the case shown in \pref{fig:m2-dist}(f), where the lattice parameters were interpolated between the two limiting cases, the M2 structure becomes energetically unstable within the R unit cell, yielding a qualitatively different picture depending on the lattice parameters used. Nevertheless, in both cases, we obtain a lower energy for the M1 structure compared to the M2 structure, consistent with the M1 phase as the global zero temperature ground state of the system. 

\subsection{Dimerization versus zigzag distortion}
\label{subsec:dimzigzag}

\begin{figure}
	\centering
	\includegraphics[width=1\linewidth]{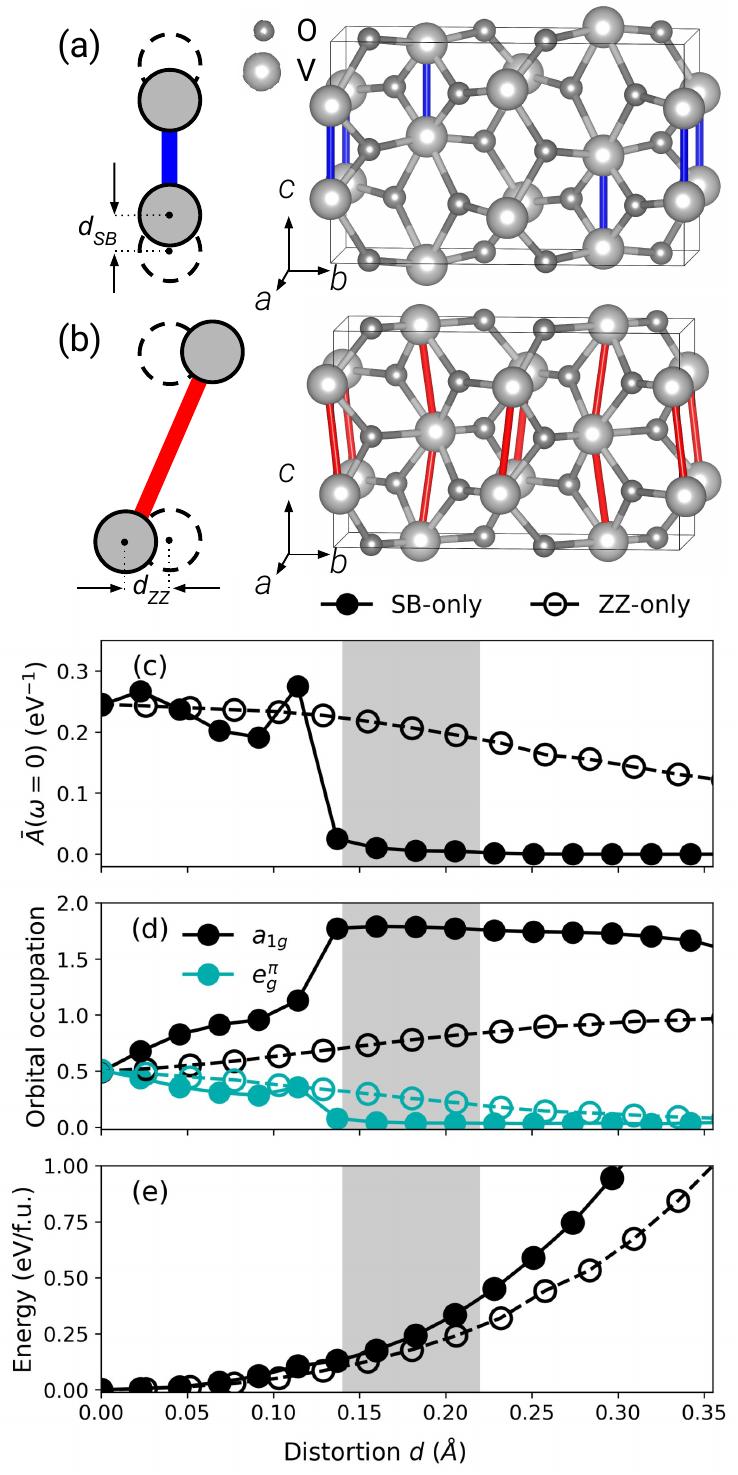}
	\caption{ (a) V--V pair undergoing a pure dimerization distortion, $d_\text{SB}$, and the resulting hypothetical SB-only structure. (b) V--V pair undergoing a pure zigzag distortion, $d_\text{ZZ}$, and the resulting hypothetical ZZ-only structure. ZZ pairs shown in red, SB pairs shown in blue. V (O) atoms shown in (dark) gray. (c-e) Selected observables calculated for SB-only (solid line, filled circles) and ZZ-only (dashed line, empty circles) structures for increasing distortion $d$ using $(U,J)=(2.0,0.1)$\,eV. The approximate range of $d$ values corresponding to the experimental M1 and M2 phases is highlighted in gray. (c) Spectral weight at zero frequency. (d) Orbital occupation of the $a_{1g}$ (black) and $e_g^\pi$ (cyan) orbitals. (e) Energy relative to the R phase.}
	\label{fig:m2-zzdimer}
\end{figure}

Next, we focus on the individual effects of the internal zigzag and dimerization distortions present in the M2 phase of VO$_2$. Here, we consider two \textit{hypothetical}, either exclusively dimerized or exclusively zigzag-distorted structures. Specifically, we displace only the V atoms either along the $c$ direction or perpendicular to it (keeping the O positions as well as the lattice parameters fixed to those of the R phase), and thus impose either a dimerization (short-bond-inducing) distortion, $d_\text{SB}$, or a zigzag distortion, $d_\text{ZZ}$, on every chain in the structure. The two resulting structures, termed ``SB-only'' and ``ZZ-only'', are shown in \pref{fig:m2-zzdimer}(a, b), highlighting the ZZ (red) and SB (blue) V--V pairs. We note that the ``SB-only'' structure of course also hosts LB sites, even though in our analysis we focus on the singlet-hosting SB sites. Furthermore, due to their construction, these hypothetical structures are more M1-like than M2-like, in the sense that they host identical distortions on all chains. They also both correspond to the same $P2_1/c$ space group as the M1 phase, which means that, since the ZZ distortion is symmetry-equivalent to the SB distortion, there are in principle two inequivalent, albeit very similar, sets of ZZ sites. Since these two types of ZZ sites do not yield any notable qualitative differences, we only report results for one type.

In \pref{fig:m2-zzdimer}(c-e), we show the evolution of the spectral weight at zero frequency, the orbital occupations, and the total energy as a function of distortion $d$ ($d_\text{SB}$ for SB-only, and $d_\text{ZZ}$ for ZZ-only), measured as the displacement of a single V atom in Å, calculated within charge self-consistent DFT+DMFT using $(U,J)=(2.0,0.1)$\,eV. We note that the M2 experimental structure hosts individual atomic displacements of $d_\text{SB} = 0.18$\,Å and $d_\text{ZZ} = 0.22$\,Å, and that the M1 experimental structure corresponds to $d_\text{SB} = 0.14$\,Å and $d_\text{ZZ} = 0.15$\,Å, resulting in a total displacement of $d=0.21$\,Å (we highlight this range of displacements in \pref{fig:m2-zzdimer}(c-e)). 

Focusing first on the SB-only structural distortion [full lines in \pref{fig:m2-zzdimer}(c-e)], we see that for increasing distortion, the system initially remains in a metallic regime, with a gradual electron redistribution into the $a_{1g}$ orbital. At $d_\text{SB} \approx 0.12$\,Å, an abrupt MIT is observed, accompanied by a sharp change in orbital occupations, resulting in a fully occupied $a_{1g}$ orbital on the short bond, i.e., similar to the MIT observed in either the M1 structure~\cite{Mlkvik_et_al:2024} or on the SB sites of M2 [\pref{fig:m2-dist}(a)]. However, unlike in these cases, we do not observe an energy lowering for the SB-only case, since the imposed distortion does not respect the elastic coupling between adjacent V chains, as previously discussed in \pref{sec:eta} and indicated in \pref{fig:m2-structure}(c), where a dimerization along one chain directly induces a zigzag distortion within the neighboring chain.

For the ZZ-only structure [dashed lines in \pref{fig:m2-zzdimer}(c-e)], we observe a rather different picture, since the system does not undergo an MIT with increasing $d_\text{ZZ}$ in the distortion range studied, even under nearly twice the experimental M1 or M2 distortion. Instead, we see a gradual decrease in the spectral weight at zero frequency, and a smooth change in orbital occupation, tending towards half-occupied $a_{1g}$ and unoccupied $e_g^\pi$ bands. This is consistent with the proposal from Ref.~\cite{Goodenough:1971} that the zigzag distortion depletes the $e_g^\pi$ orbitals, even though this depletion is in fact weaker than for the SB-only distortion. Similarly to the SB-only structure, the ZZ-only distortion mode is energetically unfavorable. The absence of an insulating state in the ZZ chains even at large distortions compared to the M2 phase indicates that the chosen value for the on-site interaction $U$ is too small to drive a Mott transition in this case.

\begin{figure}[!t]
	\centering
	\includegraphics[width=1\linewidth]{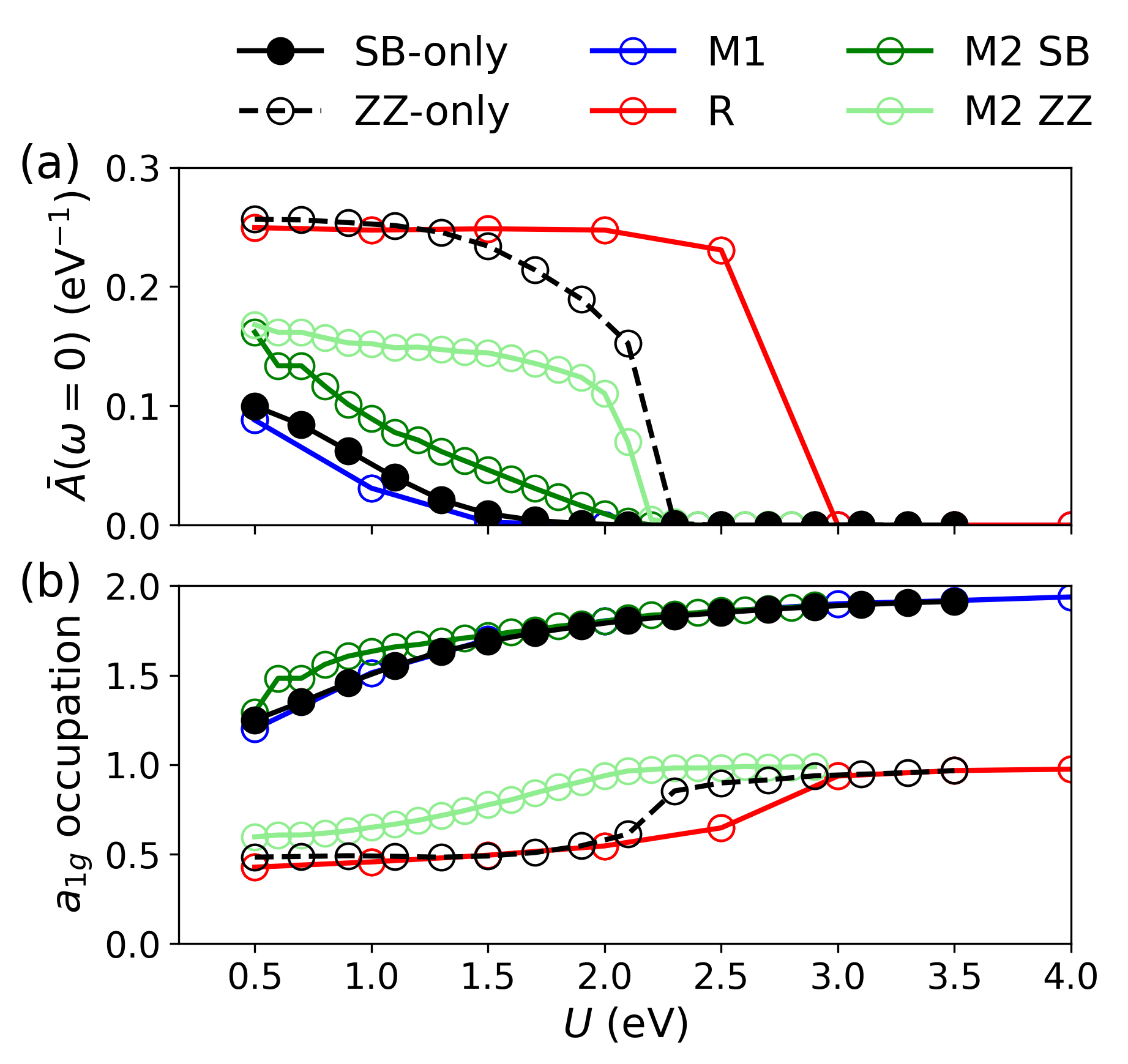}
	\caption{(a) Spectral wight at zero frequency and (b) $a_{1g}$ occupation as a function of $U$ at $J=0.1$\,eV for SB-only (black, filled circles), ZZ-only (black, dashed line), M1 (blue), and R (red) structures, and for the M2 SB site (dark green) and M2 ZZ site (light-green).}
    \label{fig:m2-zzdimer_uscan}
\end{figure}

Next, we therefore investigate the effect of $U$ on the MIT in the different structural variants, exploring the evolution of the properties as a function of the on-site interaction $U$ at $J=0.1$\,eV. This again corresponds to vertically traversing the phase diagrams in \pref{fig:m2-ujdiag}, but for different structures. In \pref{fig:m2-zzdimer_uscan}, we show the spectral weight at zero frequency and the $a_{1g}$ occupation of one site in the hypothetical SB-only (black filled markers) and ZZ-only (black empty markers) structures, both at $d=0.2$\,Å, as well as in the M1 (blue line) and R (red line) structures (data for both taken from Ref.~\cite{Mlkvik_et_al:2024}), and for the M2 SB (dark green line) and M2 ZZ (light green line) sites. All phases are calculated with one-shot DFT+DMFT in an identical cell corresponding to the experimental R lattice parameters.

Similarly to \pref{fig:m2-strain}(b, c), we observe two different types of behaviors with increasing $U$. The SB sites in the SB-only, the M2, and M1 structures undergo an MIT into a singlet insulator with an occupation of two electrons in the $a_{1g}$ orbital and a gradual decrease in spectral weight with increasing $U$, whereas the ZZ sites in the ZZ-only and M2 structures, as well as the site in the R structure exhibit a more abrupt MIT into a singly-occupied Mott insulator at some critical $U$.

Consistent with the results presented in the previous sections, the MIT occurs simultaneously at the same value of $U \approx 2.2$\,eV on both the SB and ZZ sites in the M2 structure. However, in the hypothetical SB-only and ZZ-only structures, which contain only one type of site that becomes either singlet- or Mott-insulating with increasing $U$, the corresponding MITs occur at lower or higher $U$ values, respectively, compared to the M2 case. Specifically, at around 1.7\,eV for SB-only and around 2.3\,eV for ZZ-only. Thus, the ZZ-only structure indeed becomes insulating for a sufficiently high $U$. One can further see that, compared to the completely undistorted R structure, this critical $U$ for the Mott transition is significantly reduced (from around 3\,eV for the R phase), indicating that the ZZ distortion helps the system become Mott insulating. Similarly, the M1 structure becomes insulating at a slightly lower $U$ value than the SB-only structure.

The results shown in \pref{fig:m2-zzdimer_uscan} thus suggest that in principle the transition to the singlet insulator induced by the pure dimerization distortion (as represented by $d_\text{SB}$) can occur at a lower $U$ value than the Mott transition on the undimerized ZZ chains, but that due to a cross-coupling between the transitions in the M2 phase they appear simultaneously at a $U$ value in between these limiting cases.


\section{Summary and Outlook}
\label{sec:m2-summary}

In this work, we extended the application of the bond-centered DFT+DMFT approach introduced in Ref.~\cite{Mlkvik_et_al:2024} to study the M2 phase of VO$_2$. For realistic values of the interaction parameters $U$ and $J$, we obtain a dual character of the insulating state in the M2 phase of VO$_2$, with electron singlets forming on the dimerized vanadium chains, similar to the case of the M1 phase, and a Mott-insulating character on the undimerized zigzag-distorted chains. These results are consistent with experimental observations and in good agreement with previous studies using cluster DFT+DMFT~\cite{Brito_et_al:2016, Brito_et_al:2017}, albeit obtained at much lower computational cost and without the need to pre-pattern the structure into dimerized and undimerized V--V pairs. 

Importantly, the use of bond-centered orbitals to represent the correlated subspace allows us to explore all main phases of VO$_2$, and the structural distortions connecting them, within one consistent computational framework. Our results indicate that the MIT in the M2 phase always occurs simultaneously on both the SB/LB and the ZZ chains, both as function of distortion and as function of the interaction strength, in spite of their different insulating character, and that the insulating M2 phase corresponds to a local energy minimum in the general structural phase space of VO$_2$, but is higher in energy than the M1 phase. 

Gradually distorting the structure between the energetically stable M2 and M1 structures, i.e., along a structural path corresponding to the T phase of VO$_2$, we observe a single abrupt electronic transition between an M2-like and an M1-like insulator. This appears consistent with the experimental observation of only one clear first order phase transition between the M2 and T phase, and only a continuous crossover between T and M1~\cite{Huffman_et_al:2017, Strelcov_et_al:2012}, and could suggest that the T phase can essentially be viewed as a distorted version of the M1 phase. However, further structural and electronic characterization of these phases, and the transitions between them, is needed to validate this. 

By disentangling the different components of the M2 structural distortion, i.e, the deformation of the unit cell, the dimerization, and the zigzag distortion, we showed that the unit cell strain relative to the R phase is crucial for the energetic stabilization of the M2 phase as an energy minimum (this is not the case for the M1 phase, see Ref.~\cite{Mlkvik_et_al:2024}). Furthermore, by considering the hypothetical SB-only and ZZ-only structures, we showed that both the SB-only and ZZ-only structures become insulating at different distortion and $U$ values, whereas in the M2 phase the two different insulator types become coupled and always appear simultaneously.

In summary, our computational analysis of the physical properties as function of the general structural distortion in VO$_2$, including the M2 and T phases, provides valuable insights into the long-standing problem of the coupled electronic and structural transition in VO$_2$, and highlights the importance of treating the entire structural phase space using one consistent approach that allows to specifically study also the transitions between the different observed phases. Since the bond-centered DFT+DMFT approach treats each V--V bond on equal footing and does not involve a pre-patterning of the underlying structure, it is particularly promising to also assess the effects of, e.g., applied strain or defects such as oxygen vacancies on the stability between different phases and their electronic properties. Such future studies can potentially provide valuable insights on how to optimize the MIT in VO$_2$ for different anticipated technological applications.


\section*{Acknowledgments}
This work was supported by ETH Z\"{u}rich and the Swiss National Science Foundation (Grant No.~209454). Calculations were performed on the ETH Z\"{u}rich Euler cluster and the Swiss National Supercomputing Center Eiger cluster under Project ID No. s1304.

\section*{Data availability}

The supporting data for this article are openly available on the Materials Cloud Archive~\cite{zotero-item}. 

\bibliography{M2}

\end{document}